\documentclass[aps,prb,reprint,showpacs,superscriptaddress,amsmath,amssymb]{revtex4-1}
\usepackage[utf8]{inputenc}

\usepackage{graphicx}
\DeclareGraphicsExtensions{.eps}

\usepackage{bm}
\usepackage{textcomp} % This package is just to give the text quote '
\usepackage{enumitem}
\usepackage{amsmath,amssymb}
\usepackage{graphicx}
\usepackage{color}
\usepackage{xcolor}
\usepackage{amsmath}
\usepackage{blindtext, rotating}
\setlist{noitemsep,leftmargin=*}

\usepackage{hyperref}
\hypersetup{
	colorlinks=true,
	urlcolor= blue,
	citecolor=blue,
	linkcolor= blue,
	bookmarksopen=false,
}

\begin{document}

\title{Phonons in  MoSe$_2$/WSe$_2$ van der Waals heterobilayer}

\author{F. Mahrouche}
\affiliation{Laboratoire de Physique Th\'eorique, Facult\'e des sciences exactes, Universit\'e de Bejaia, 06000 Bejaia, Alg\'erie}
\affiliation{Departamento de F\'{i}sica Aplicada, Universidad de Alicante, 03690 San Vicente del Raspeig, Spain}

\author{K. Rezouali}
\affiliation{Laboratoire de Physique Th\'eorique, Facult\'e des sciences exactes, Universit\'e de Bejaia, 06000 Bejaia, Alg\'erie}

\author{F. Zaabar}
\affiliation{Laboratoire de Physique Th\'eorique, Facult\'e des sciences exactes, Universit\'e de Bejaia, 06000 Bejaia, Alg\'erie}

\author{A. Molina-S\'{a}nchez}
\affiliation{Institute of Materials Science (ICMUV), University of Valencia, Catedr\'{a}tico Beltr\'{a}n 2, E-46980, Valencia, Spain}

\begin{abstract}

We report first-principles calculations of the structural and vibrational properties of the synthesized two-dimensional van der Waals heterostructures formed by single-layers dichalcogenides MoSe$_2$ and WSe$_2$.  We show that, when combining these systems in a periodic two-dimensional heterostructures, the intrinsic phonon characteristics of the free-standing constituents are to a large extent preserved but, furthermore,   exhibit shear and breathing phonon modes that are not present in the individual building blocks. These peculiar modes depend strongly on the weak vdW forces and has a great contibution to  the thermal properties of the layered materials. Besides these features, the departure of flexural modes of heterobilayer from the ones of its monolayer parents are also found.

\end{abstract}
\maketitle
\section{Introduction}
Comprised of one or several covalently bound sheets of atoms held together by van der Waals (vdW) forces, two-dimensional (2D) systems and layered weakly-bound structures thereof \cite{Novoselov666,Novoselov10451,Coleman568}   have interesting properties that make them highly desirable in many fields. Besides their
fascinating properties, they have the potential to sharply reduce the characteristic lengths of electronic devices thus opening up unprecedented perspectives in view of designing low-power devices such as ultra-thin-channel field-effect transistors and solar cells.
The family of transition metal dichalcogenides (TMDs) is one of these engaging materials, consisting of  atomically thin semiconductors of the type MX$_{2}$ where M is a transition metal atom (Mo or W, etc.) and X is a chalcogen atom (S, Se or Te).
They exhibit small direct bandgaps  lying in visible and near-IR ranges \cite{Xia2014} making them  attractive shortly after they were isolated due to their unique physicochemical properties and potential applications in electronic and optoelectronic devices. 
For example, WSe$_2$ is used as field-effect transistors (FETs) \cite{Fang2012}, logic-circuit integrations , and optoelectronic devices \cite{huang_large-area_2014}, while MoSe$_2$ is used in field effect transistors \cite{Larentis2012}, solar cells and photoelectrochemical devices \cite{Tongay2012, Jariwala2014, Krasnok2018} . \\

Beyond 2D materials considered individually, composite systems stacks of different layers, called van der Waals
heterostructures \cite{Geim2013} are particularly attractive: They allow not only to modulate the properties of the materials, but also to further improve their performance.\\
Recently, many efforts are devoted to investigate the different properties of a wealth of vdW hybrid heterostructures (vertical and lateral) such as MoSe$_{2}$/WSe$_{2}$ \cite{Gong2015},WSe$_{2}$/MoS$_{2}$ \cite{Lee2014} MoS$_{2}$ /h-BN/graphene \cite{Lee2013}.\\
%The vdWHs of transition metal dichalcogenides (TMDs)are of particular interest for optoelectronic and electronic technologies.
In particular,  many  experimental and theoretical studies of the physical properties  of MoSe$_{2}$/WSe$_{2}$ heterostructure have been done because of  their potential  for optoelectronic and electronic technologies and light harvesting applications. \cite{Hanbicki2018,Rigosi2015,Miller2017,Gillen2018,Torun2018,Zhu2017}
Despite the effort devoted the study of the dynamical and vibrational properties of the heterostructures based on the  MoSe$_{2}$ and WSe$_{2}$ are scarse. 

In this work, we investigate the structural and vibrational properties of MoSe$_{2}$/WSe$_{2}$ van der Waals heterostructures with a special focus
on the observed interlayer phonon modes in the studied systems. The phonon characteristics of the heterobilayer (HBL) are calculated in the framework of lattice dynamics theory using appropriate ab-initio methods. 
These present-day theories can compute phonons in many systems without any adjustable parameters making them complementary tools to the measurements.
We explore emergent properties in the composite system that are absent in the individual layers. Our calculation reveals the existence of  interlayer shear and breathing phonon modes in the heterobilayers. These peculiar phonons exist only in bilayer areas and are paramount to understand the different scattering mechanisms in layered 2D materials.
 
The manuscript is organized as follows.
Descriptions of the computational methods are given in
Section II. In Section III, we report and discuss our results,
and finally we summarize our main conclusions.

\section{Methods}

\begin{figure}[h]
	\centering
	\includegraphics[scale=0.04]{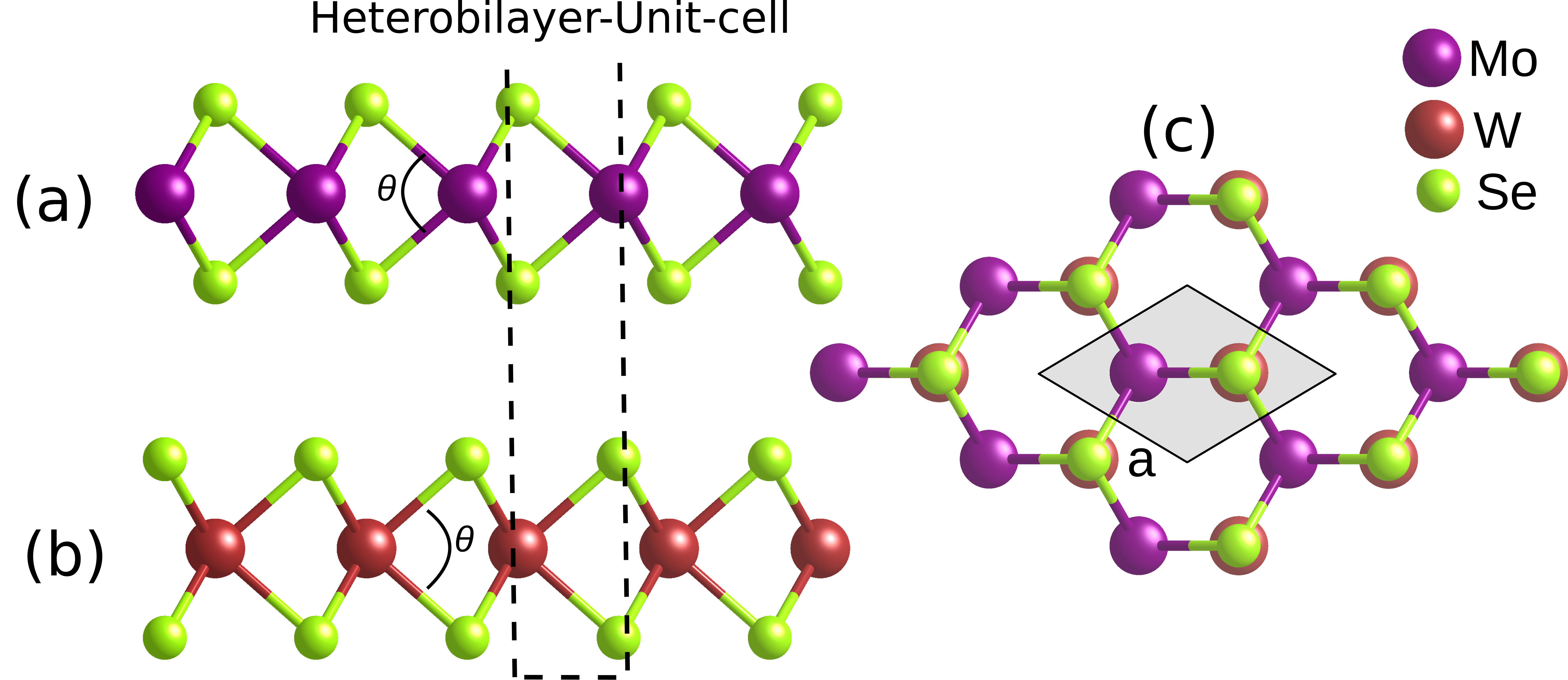}
	\caption{(Color online) Side view and top view of atomic structures of van der Waals heterobilayer MoSe$_{2}$/WSe$_{2}$ .}
	\label{structure}
\end{figure}

XSe$_{2}$ \{ X=Mo or W \} are  transition metal dichalcogenides consisting of covalently bound Se-X-Se sheets that are held together by van der Waals (vdW) forces as shown in Fig.\ref{structure}. Each one of these single layers consists of hexagonal plane of metallic X atoms sandwitched between two hexagonal planes of Se atoms  bound with the metallic atoms in a trigonal-prism arrangement.

The unit cell is constructed by combining a 1 $\times$ 1 unit cell of monolayer MoSe$_{2}$  and 1 $\times$ 1 unit cell of WSe$_{2}$ sheet. The primitive lattice vectors spanning the heterobilayer supercell are chosen in commensuration with those of the isolated layers. The supercell containing 6 atoms is depicted in Fig.\ref{structure}. The determination of  the equilibrium geometry of any structure constitutes the pivotal step before investigating phonons or any other related physical properties. The calculations are carried out using density functional theory (DFT) \cite{Hohenberg1964, PhysRev.140.A1133} as implemented in the open-source code  QUANTUM ESPRESSO.\cite{Giannozzi2009}  We adopted the Perdew-Burke-Ernzerhof exchange correlation functional. \cite{Perdew1996} The ionic potential has been modeled using Optimized Norm-Conserving Vanderbilt (ONCV) pseudopotentials.\cite{hamann_optimized_2013}  
To correctly describe the effect of dispersive forces, we employed the Tkatchenko-Scheffler (TS) dispersion corrections method \cite{Tkatchenko2009}  which has been demonstrated to give a reliable prediction for both interlayer distances and binding energies for vdW systems. The calculations were done in the supercell arrangement using a 80 Rydberg (Ry) plane-wave cutoff for
the electronic wave function expansion. To maintain the periodicty along the direction perpendicular to the sheet, we used a vacuum  of 18 \AA. The first Brillouin zone is sampled with a 12 $\times$ 12 $\times$ 1 Monkhorst-Pack grid \cite{Pack1977} together with a simple Gaussian smearing in the electronic occupations of 0.02 Ry for single-layer and heterobilayer systems. 
This technique is also used to avoid the spurious interlayer interactions. The equilibrium atomic coordinates has been obtained fully minimizing unit cells using the calculated forces and stress on the atoms.
The convergence criterion of self-consistent calculations for ionic relaxations is 10$^{-10}$ eV  between two consecutive steps. All atomic positions and unit cells are optimized until the atomic force convergence of 10$^{-3}$ eV/\AA ~ were reached. The above criteria ensure the absolute value of stress is less than 0.01 kbar.

To benchmark our structural calculations, the lattice parameters, bond lengths, and bond angles of monolayers WSe$_{2}$ and MoSe$_{2}$ are determined beforehand and gathered in Table \ref{T1}.  The calculated lattice constants are evaluated to be $a$ = 3.31 \AA ~  for both monolayers MoSe$_{2}$ and WSe$_{2}$.  The Mo$-$Se and W$-$Se bond lengths are calculated to be about 2.54 \AA ~ and 2.55 \AA ~, respectively. The out-of-plane  Se$-$Se bond lengths is evaluated to be 3.34 \AA ~ and 3.39 \AA ~ for MoSe$_{2}$ and WSe$_{2}$, respectively.  
The out-of-plane $ \angle$ SeMoSe and  $\angle$ SeWSe bond angles are calculated to be about 82.19\textdegree ~  and  83.07\textdegree, respectively. Our results are in accordance with available theoretical
and experimental data.\cite{Ding2011,Chang2013,Ramasubramaniam2012}.
For  heterobilayer, the lattice constant $a$ is decreased when compared with that of isolated layers by about 0.6 \%. The Mo$-$Se and W$-$Se bond lengths are calculated to be about 2.53 \AA ~ which decreases by less than 0.8 \% than those of its parent layers reflecting the weak interlayer interaction. The interlayer distance $d$ between MoSe$_{2}$ and  WSe$_{2}$ monolayers is 3.22 \AA. It should be noted that our structural forecasts are in good agreement with the theoretical data. \cite{Debbichi2014}

The phonon characteristics of a given material are often described in the context of lattice dynamics theory by using appropriate first-priciples methods. The inputs  are the second interatomic force constants (IFCs). IFCs  were calculated using density functional perturbation theory (DFPT).\cite{RevModPhys.73.515}  Non-analytical terms due to the Coulomb forces, Born charges, and dielectric constants are all gathered in the dynamical matrices for both the subunits and their descending heterobilayer. Phonon group velocity is calculated based on the phonon dispersion relation for each single mode.
For the calculation of the harmonic IFCs, a Wigner-Seitz (primitive)  equilibrium cell of each structure of interest is used.   
The harmonic IFCs were performed on 12 $\times$ 12 $\times$ 1  \textbf{k} point meshes for phonon calculations of single-layers and heterobilayer, with a simple Gaussian smearing in the electronic occupations of 0.02 Ry. The above samplings are sufficient to map out all the possible instabilities and offer well-converged phonon eigenvalues for all of the structures treated here.
\begin{table}[h]
	\resizebox{0.5\textwidth}{!}{
		\begin{tabular}{l|llllllll}

			Material		& $a$  & ~$d_{X-Se}$ ~~~ $d_{Se-Se}$~~~  $d$ ~~  $\angle$ SeMoSe & ~~ $\angle$ SeWSe &\\ 
			\hline\hline\\ 
			MoSe$_{2}$ 	& 3.31  & ~~ 2.54 ~~ ~ 3.34 ~~~~~~~~~~~~~~~ 82.19  & ~~~~~~~   \\
			
			\hline\\

			WSe$_{2}$   & 3.31 & ~~ 2.55 ~~ ~ 3.39 ~~~~~~~~~~~~~~~   & ~~~~~~ 83.07 \\
			\hline\\
			MoSe$_{2}$/WSe$_{2}$  & 3.29 & ~~  2.53  ~ 3.35/3.37 ~~~ 3.22 ~~ 82.55 & ~~~~~~ 82.84 & \\
			\hline\hline                             
	\end{tabular}}
	
	\caption{Calculated values of $ MX_{2}$  single layer 2D, lattice constants,	$a$, bond lengths, $d_{M-X}$ and $d_{X-X}$  in angstrom (\AA), and bond angles in degree. } 
	\label{T1}
\end{table}

Because MoSe$_{2}$ and WSe$_{2}$ are polar materials, nonanalytic part of the dynamical matrix which contains the effective charges and the dielectric tensor are taken into account to obtain the correct frequencies at the zone center. Indeed, certain IR-active mode (polar mode) at the zone-center give rise to a macroscopic electric field which impacts the longitudinal optical (LO)
phonons in the limit of long-wavelength, breaking the degeneracy of the LO mode with the transversal optical (TO) mode.\cite{Cardona}  The LO-TO splitting for the $E^{'}$ arises from the coupling of the lattice to the macroscopic electric field created by the relative displacement of Mo ,W atoms and Se atoms in the long-wavelength limit. \cite{Wirtz2011,Cai2014}
\section{Phonon dispersion}
Phonons play a key role in a wide range of phenomena in condensed matter physics. For instance the major contributions of
lattice thermal conductivity  \cite{ Zhan-yu2018} and electron-phonon coupling depend on phonons. \cite{Jin2016, Nam2016} Recently several studies have been devoted to studying the vibrational properties of graphene \cite{Yan2008,Wang2009,Chun2013},  graphene/TMDs heterostructures\cite{Singh2018,C6CP02424F}, transition metal dichalcogenides (TMDs) \cite{Ding2011,Wirtz2011,Zhao2013,Luo2013,Tornatzky2018, Yanyuan2013,Nam2016,Lui2015,Sandoval1991}, and their stacked heterostructures.\cite{Nam2016, Lui2015} The observation of arising interlayer phonon modes \cite{Chun2013,Yanyuan2013,Nam2016,Lui2015} in van der Waals heterostructures  is relevant because describe the interaction between layers. Here, we explore MoSe$_{2}$/WSe$_{2}$ vdW heterobilayer, focusing on the way the intrinsic vibrational properties of the building blocks are modified upon stacking.

 \begin{table}
 	\resizebox{0.5\textwidth}{!}{
 		\begin{tabular}{ccccccccccc}
 			\hline 
 			\hline 
 			Mode & Activity &   &  Wse2 &   &       &  & MoSe2 &  \\
 			\hline
 			&     & Cal     & Ref\cite{Li2014}  &Ref\cite{Ding2011}   & Exp\cite{Zhao2013}   &Cal   &Ref\cite{Li2014}&  Ref \cite{Ding2011}& Exp\cite{Tongay2014}\\  \\ \hline
 			$E''$& R          &164.9 & 168.11& 166 &  -   & 161.8  &161.77 & 160 &- \\ \\
 			$A'_{1} $ & R       &236.6& 241.50& 241  & 248   & 234.2  &236.16 & 233 & 241\\ \\ 	
 			
 			$E'$	 & I+R       & 234.7(234.8) & 238.49& 238  & 249    & 276.9(279.8) & 277.85 &277 & 287 \\ \\
 			
 			$A''_{2}$& I         &293.5 & 298.53& 298  &  -    &  342.5 &342.90 &343  &-  \\ \\  \hline
 		
 	\end{tabular}}
 	\caption{Zone-center phonon modes of  monolayers MoSe$_{2}$ and WSe$_{2}$.   } 
 	\label{Ta2}
 	
 \end{table}

\subsection{  Individual Layers}

\begin{figure}[h]
	\includegraphics[scale=0.15]{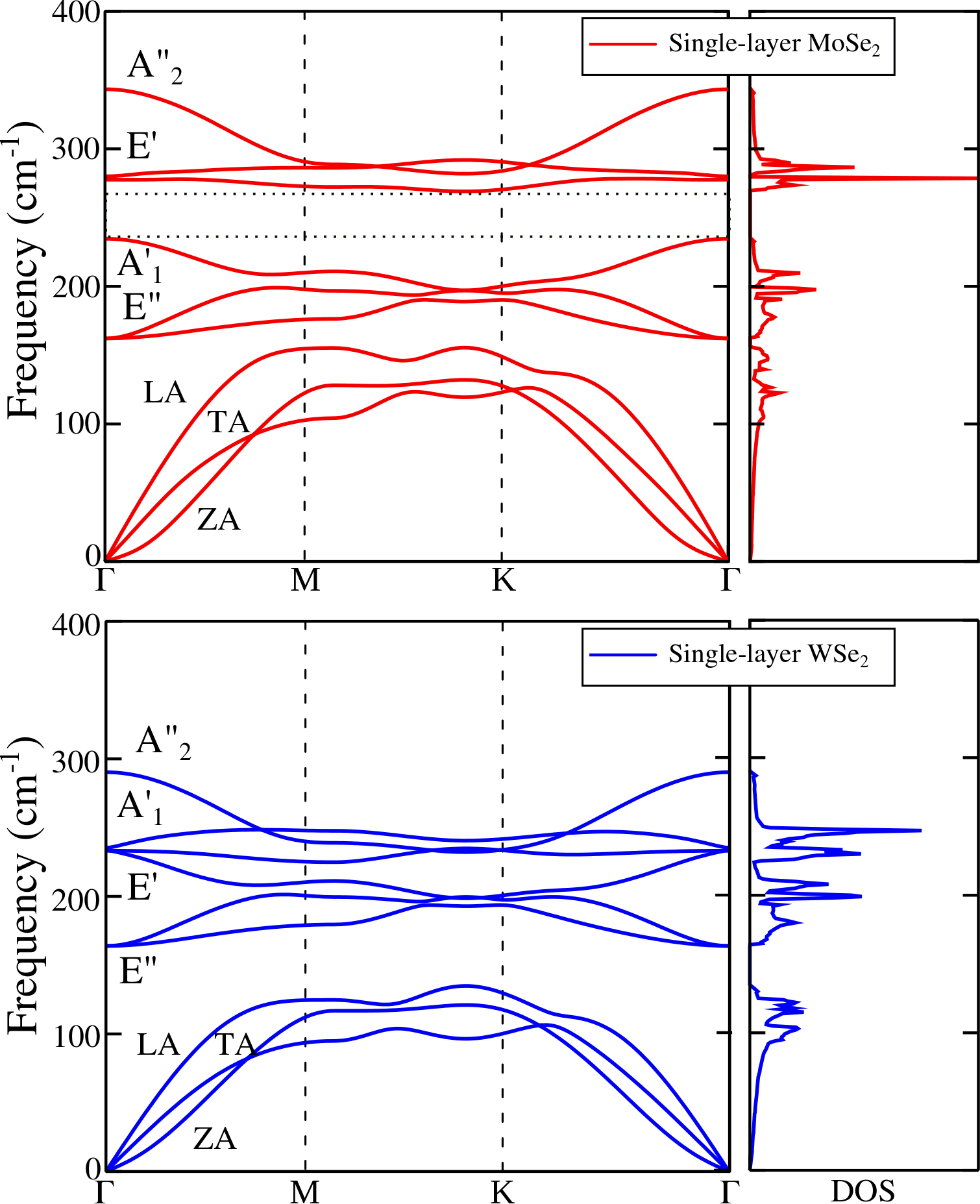}
	\caption{(Color online) Phonon dispersion and phonon DOS for monolayer  MoSe$_{2}$  and WSe$_{2}$.}
	\label{phonon}
\end{figure}

The phonon dispersions at zero stress along the main symmetry directions and its density of states (DOS) are shown in Figs.\ref{phonon} (a),(b). Overall, the MoSe$_{2}$ and WSe$_{2}$ phonon dispersions share a resemblance. It is worth noting the absence of degeneracies at the high-symmetry points M and K and the  crossing of the LA and ZA branches just before the M point. 
The most relevant phonon frequencies at $\Gamma$ with their character, involved atomes, and displacement direction  are summarized in Table \ref{Ta2}. We shall emphasize the good agreement between our calculated dispersion and those available in the literature\cite{Johari, Hu2016i}.

\begin{figure}[h]
	\includegraphics[scale=0.25]{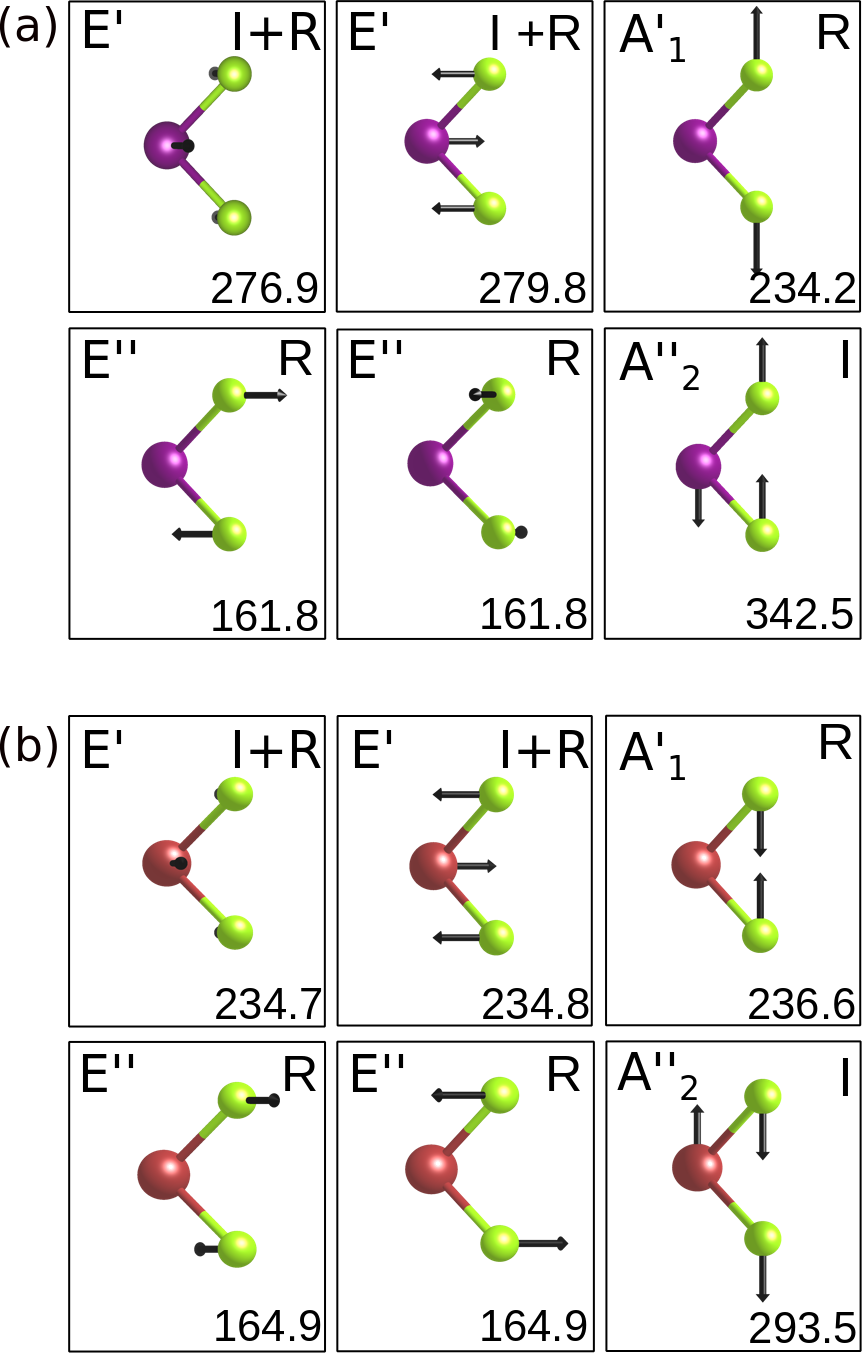}
	\caption{(Color online) Atomic displacements for optical phonon modes of monolayers  MoSe$_{2}$ (a) and (b) WSe$_{2}$.} 
	\label{ph}
\end{figure}

At the $\Gamma$  point both MoSe$_{2}$  and WSe$_{2}$ possess the factor group $P$-6$m$2 (D$_{3h}$ in Sch\"onfliess notation) with no center of inversion. The zone-center vibrational modes can be decomposed according to the following equation  \cite{Sandoval1991,Almeida2014}

\begin{center}
	$\Gamma= A_2^{''}\oplus 2E^{'}\oplus2E^{''}\oplus A_1^{'} \oplus 2E^{'}\oplus A_2^{''}$
\end{center}

The superscripts $'$ and $''$ represent modes that are symmetric or anti-symmetric  with respect to the horizontal plane $\sigma_{h}$. The subscripts 1 and 2 refer to modes that are symmetric or anti-symmetric to $C_{ 2}$ perpendicular to principal axis $C_{ n}$  or vertical plane  $\sigma_{v}$ or  $\sigma_{d}$. Furthermore,  the symbol $E$ represents the doubly degenerate vibratory modes while the symbol A stands for the non-degenerate vibratory modes.

The three acoustic modes are either IR-active or Raman active and can be decomposed into irreducible representations
\begin{center}
	$\Gamma^{acoustic}= A_2^{''}\oplus 2E^{'}$
\end{center}
A degenerate $E^{'}$ mode which is devided into in-plane transversal $E^{'}$ TA and longitudinal $E^{'}$ LA mode that are both Raman and IR-active involves in-plane motions of atoms. Both LA and TA modes show linear dispersion in the long wavelength limit.  The calculated LA and TA sound velocities are extracted from the slope of the acoustic branches. We obtain 4.71  (4.33) km/s and 2.91 (2.77) km/s for the LA and the in- plane TA branches, respectively in the $\Gamma-$M direction. The values between round brackets stand for  WSe$_{2}$.
In the direction $\Gamma-$K both values  are slighly lower by about  1.4\% (0.7\% ) for TA phonon branch and 7\% (1.9 \%) for LA phonon branch than the ones obtained in the direction $\Gamma-$M. Such a direction dependency of the phonon propagation velocity has not been reported so far. It may have a big impact on physical properties that are related the sound velocity such as the elastic constants, in-plane stiffness, and shear modulus.

The mode $A_2^{''}$ that is IR-active belongs to the out-of-plane mode ZA where the atoms move out of the plane. ZA phonons  are
nonlinear with respect to the wave vector with an approximate quadratic $q^2$  dependence as a consequence of the groupe symmetry. These flexural modes are particularly important for low-dimensional layered systems \cite{Samuels,Jiang_2015}   as they have a huge contribution to the phonon DOS and are responsible for the large thermal conductance. It was also reported that the transport due to the flexural excitations is almost ballistic.\cite{Nam2016} 
Fitting the frequencies below 40 cm$^{-1}$ to a parabolic function $\omega $ = $\delta $ $q^2$ , a value of $\delta $ $\approx$ 5.667 $\times$ 10$^{-7}$  $m^2$ $s^{-1}$ and 5.648 $\times$ 10$^{-7}$  $m^2$ $s^{-1}$ is obtained for MoSe$_{2}$ and WSe$_{2}$, respectively, which values are comparable  to that of graphene ($\approx$ 6 $\times$ 10$^{-7}$) \cite{daniel1998,LUCAS1993587}.

Finally, using the $q^2$  behaviour of the ZA branch we have estimated the energy necessary to roll up the MoSe$_{2}$ and WSe$_{2}$
sheet to form the tubules. To do so, we used the equation of strain energy per X$-$Se atom pair $E_{st}$ = $A/r^2$ where $A =\delta^2 (m_X +m_{Se})/2$, being ($m_X +m_{Se}$) the mass of the X$-$Se pair and $r$ the tube radius. The value obtained for A is 5.13 eV \AA$^2$ /pair and  7.86 eV \AA$^2$ /pair for MoSe$_{2}$ and WSe$_{2}$, respectively , which are in the same magnitude as compared to the calculated constant for graphene and carbone nanotubes \cite{daniel1998}.

The optical zone-center modes are decomposed into 
\begin{center}
$\Gamma^{optical}=2E^{''}\oplus A_1^{'} \oplus 2E^{'}\oplus A_2^{''}$
\end{center}

The eigenvectors and frequencies for the optically allowed $\Gamma$-point vibrations are shown in Fig.3. The doubly degerate $E^{"}$ Raman-active modes LO$_1$ and TO$_1$ at 161.8 cm$^{-1}$ (164.9 cm$^{-1}$) involve rigid shear diplacements  between two neighboring Se layers in the primitive cell while the metallic atoms remain frozen. The nondegenerate out-of-plane $A^{'}_{1}$  mode ZO$_2$ at 234.2 cm$^{-1}$ (236.6 cm$^{-1}$) that is  Raman active consists of an out-of-plane stretching for MoSe$_{2}$ and of a rigide plane compression for WSe$_{2}$.

The high-frequency optical modes are separated from the low-frequency modes by a gap of 42.6 cm$^{-1}$.
The doubly degenerate vibratory $E^{'}$ mode splits into LO$_2$ mode at  276.9 cm$^{-1}$   and TO$_2$ at 279.8 cm$^{-1}$ that are both Raman and IR-infrared and involve rigid-plane shear diplacements between  the metallic layer and the chalcogenide layers in the primitive cell. In the case of WSe$_{2}$ single-layer the LO-TO splitting for the $E^{'}$ mode has the value of 0.1 cm$^{-1}$ which is much smaller than in the MoSe$_{2}$ single-layer.

 The nondegenerate out-of-plane $A^{''}_{2}$  mode ZO$_1$ at 342.5 cm$^{-1}$ and 293.5 cm$^{-1}$ for MoSe$_{2}$ and WSe$_{2}$, respectively, that is IR-active consists of an out-of-plane stretching of the metallic atoms and chalcogenide atoms. It is worth mentioning the good agreement between the calculated phonon frequencies of single-layers MoSe$_{2}$ and WSe$_{2}$ and the experimental and theoretical results ( see Table \ref{Ta2}).

\subsection{ MoSe$_{2}$/WSe$_{2}$ heterobilayer}
 Figure 5 displays the phonon dispersions along the main symmetry directions and the phonon DOS for MoSe$_{2}$ /WSe$_{2}$ heterobilayer.
 Although that the phonon dispersion curve looks similar to those of the subunits, indicating the low effect of  vertical stacking on phonon structures, the  phonon structure of the heterobilayer is not a simple superposition of those of isolated layers. The unit cell of the heterostructure considered here contains six atoms, resulting in eighteen phonon branches. 
 At the $\Gamma$ point they possess the factor group C$_{3v}$ (3m). The zone-center vibrational modes can be decomposed into the following irreducible representations 
 \begin{center}
 	$ \Gamma^{HBL} = 6A_{1}\oplus 12E$ 
\end{center}
All the zone-center phonons modes are both Raman and infrared active.
The three acoustic modes can be decomposed into $\Gamma^{acoustic} = A_{1}\oplus 2E$ irreducible representations. In a similar way, these three branches constitute the acoustic modes ZA, TA and LA as it is quoted before in the monolayers acoustic branches. 

From Fig. 5 we can see that similar to other 2D hexagonal materials \cite{C5RA19747C,Pokatilov,Chiencheng}, the longitudinal acoustic (LA) and transverse acoustic (TA) branches show linear dependence in the vicinity of the $\Gamma$ point. Notice that no  imaginary frequencies occur along $\Gamma$ -M  and  $\Gamma$ - K  line directions for out-of-plane acoustic ZA branch indicating that the lattice is thermodynamically stable even for long-wavelength transverse thermal vibrations. The speed of sound  is calculated to be about 4.79  km/s and 4.74  km/s for the LA branch  in the $\Gamma-$M direction and $\Gamma-$K direction, respectively. For in- plane TA branches, the value for the sound velocity is 3.02 km/s for both $\Gamma-$M and $\Gamma-$K directions. It is worth to mention that the sound waves travel more faster through heterobilayer than in the isolated constituents. The explanation for this trend is the compression waves (longitudinal waves) that travel fastest. This type of body waves does not exist in the isolated layers.

Examination of the dispersion curve shows that the flexural modes have departed from the quadratic wave vector dispersion and the frequency is $\omega $ = $\delta $ $q^x$, where $x$ =1.44,  which value is comparable  to that of graphene/MoS${_2}$ HBL (1.45).\cite{Nam2016} The departure from the $q^2$ behavior is an indication that the heterobilayer is less elastic than the free-standing layers.\cite{Chun2013}
Fitting the frequencies to a function $\omega $ = $\delta $ $q^{1.44}$ , a value of $\delta $ $\approx$ 1.977 $\times$ 10$^{-7}$  $m^2$ $s^{-1}$ is obtained for MoSe$_{2}$ / WSe$_{2}$ heterobilayer,  which values are lower than that of isolated layers confirming again the transition from an elastic systems to less elastic system. 

Besides the three acoustic modes, there are 15 optical zone-center modes that are decomposed into 

\begin{center}
$\Gamma^{optical} = 5A_{1}\oplus 10E$.
\end{center}

The  optical phonon frequencies at $\Gamma$ point for the  heterobilayer are summarized in Table \ref{tab3}. As it can be seen, the intrinsic phonon characteristics of the free-standing constituents are to a large extent preserved but, furthermore, it exhibits peculiar phonon modes that are not present in the individual building blocks. In other words, the stacking arrangements give rise to two types of optical phonons modes: ten subunits-like phonon (five MoSe$_2$-like phonons and five WSe$_2$-like phonons), and five hybrid-like phonons shown in Fig.6.

\begin{figure}[h]
	\includegraphics[scale=0.1]{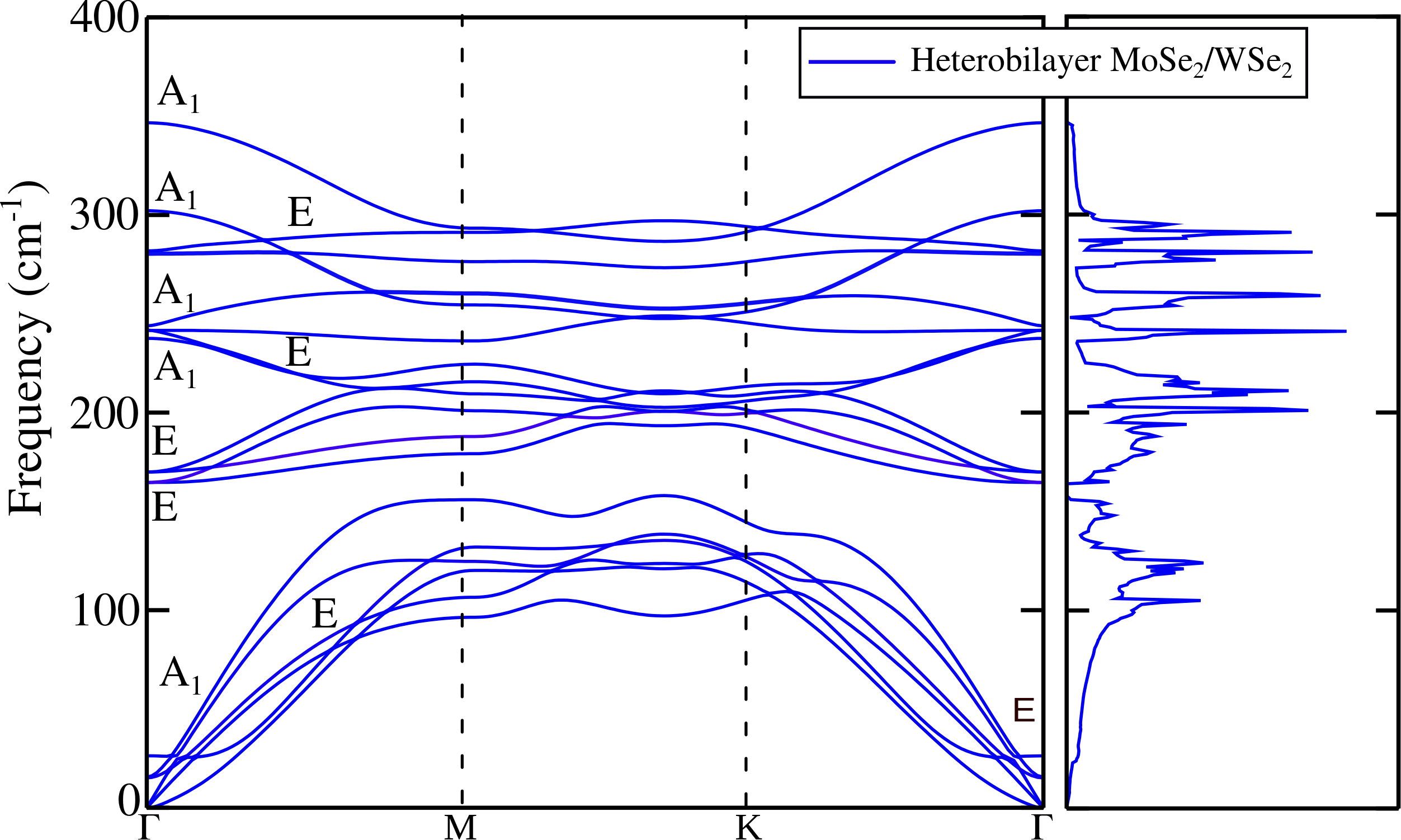}
	\caption{(Color online) Phonon dispersion and phonon DOS of heterobilayer MoSe$_{2}$/WSe$_{2}$.}
	\label{ph}
\end{figure}

\begin{table}[h!]
\begin{tabular}{cccccc}
		\hline \hline$C_{3 v}$ & Character & Type & Direction & Atoms & $\omega$(cm$^{-1}$)\\
	\hline$E$ & $\mathrm{R}+\mathrm{I}$ & Interlayer & $/ /$ & $\mathrm{Mo}+\mathrm{W}+\mathrm{Se}$ & 15.0 \\
	\hline$A_{1}$ & $\mathrm{R}+\mathrm{I}$ & Interlayer & $\perp$ & $\mathrm{Mo}+\mathrm{W}+\mathrm{Se}$ & 27.0 \\
	\hline$E$ & $\mathrm{R}+\mathrm{I}$ & MoSe$_{2}$-like & $/ /$ & $\mathrm{Se}$ & 164.3 \\
	\hline$E$ & $\mathrm{R}+\mathrm{I}$ & WSe$_{2}$-like & $/ /$ & $\mathrm{Se}$ & 170.4 \\
	\hline$A_{1}$ & $\mathrm{R}+\mathrm{I}$ & Interlayer  & $\perp$ & $\mathrm{Se}$ & 237.4 \\
	\hline$E$ & $\mathrm{R}+\mathrm{I}$ & WSe$_{2}$-like & $/ /$ & $\mathrm{W}+\mathrm{Se}$ &  241.8 \\
	\hline$A_{1}$ & $\mathrm{R}+\mathrm{I}$ &  Interlayer & $/ /$ & $\mathrm{Se}$ & 244.2 \\
	\hline$E$ & $\mathrm{R}+\mathrm{I}$ & MoSe$_{2}$-like   & $\perp$ & $\mathrm{Mo}+\mathrm{Se}$ & 280.4 \\	
	\hline$A_{1}$ & $\mathrm{R}+\mathrm{I}$  &  WSe$_{2}$-like  & $/ /$ & $\mathrm{W}+\mathrm{Se}$ & 302.3 \\
	\hline$A_{1}$ & $\mathrm{R}+\mathrm{I}$ & MoSe$_{2}$-like  & $\perp$ & $\mathrm{Mo}+\mathrm{Se}$ & 346.6 \\
	\hline \hline
\end{tabular}
\caption{
Zone-center phonon frequencies and relevant phonon symmetry representations of heterobilayer MoSe$_{2}$/WSe$_{2}$ (point group $C_{3 v}$ ).  Direction $/ /$ ($\perp$) is parallel (perpendicular) to the $c$ vector of the unit cell, respectively.}
\label{tab3}
\end{table}

The hybrid modes also called interlayer modes can be devided into interlayer shear phonon modes (LSM) and breathing phonon modes (LBM). These vibrational modes describe the relative displacement of WSe$_{2}$ monolayer with respect to the MoSe$_{2}$ monolayer in in-plane and out-of-plane directions. 
\begin{figure}[h]
	\includegraphics[scale=0.07]{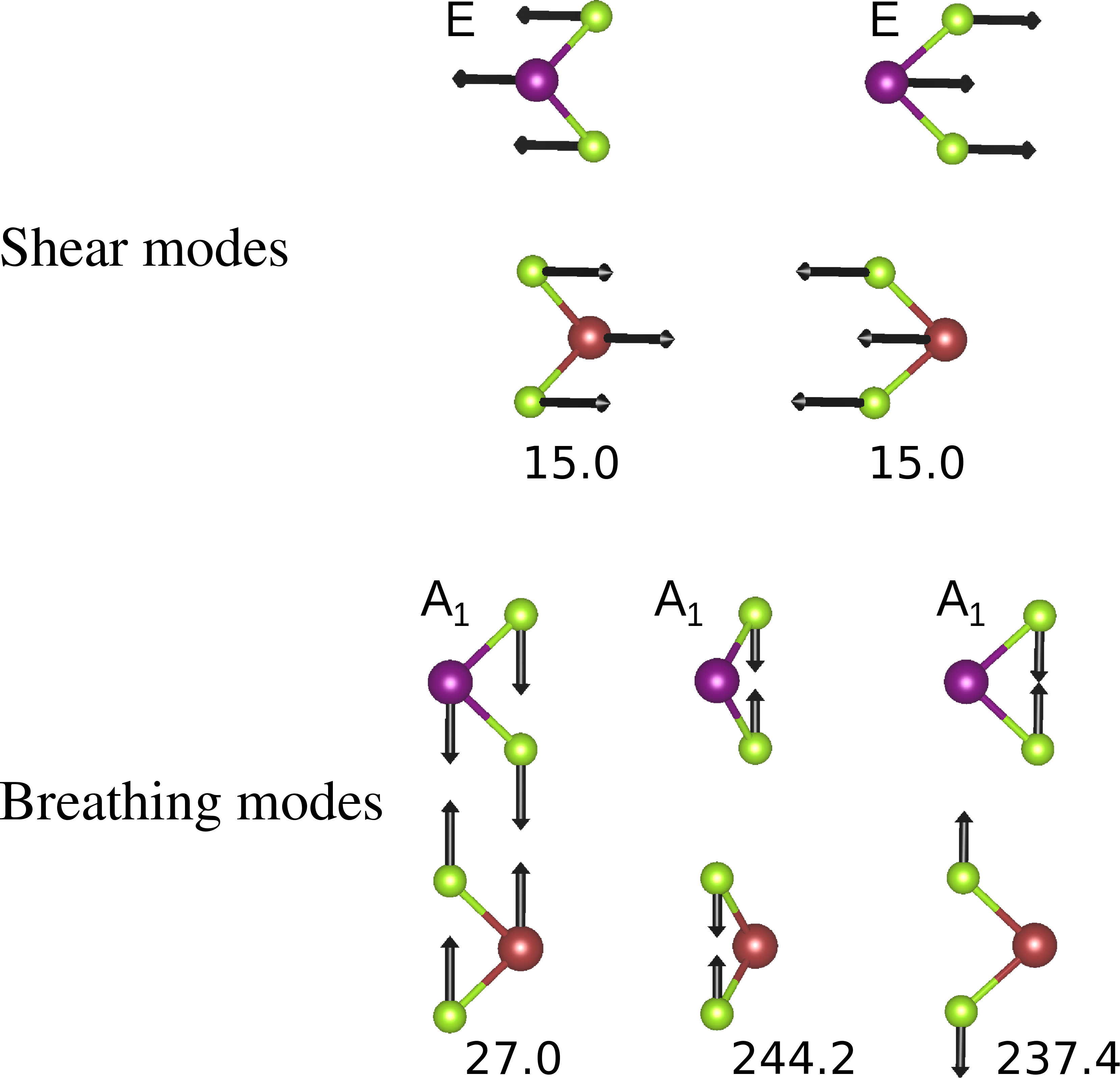}
	\caption{(Color online) Atomic displacements for optical phonon modes of heterobilayer MoSe$_{2}$/WSe$_{2}$.}
	\label{ph}
\end{figure}

Considering the symmetry properties of the cristal, we expect two shear phonon modes and three breathing phonon modes as illustrated in Fig.6. The rigid-layer shear modes (doubly degenerate) start at 15.0 cm$^{-1}$ at $\Gamma$ point. It is worth to notice that this  mode vibration cannot be generated in the heterobilayers with incommensurate lattices because the lateral displacement of the two layers are not capable of producing any overall restoring force.\cite{Lui2015}
The frequencies of the three breathing phonon modes at the zone-center are 27, 237.4 and 244.2 cm$^{-1}$. Note that when the wave number $\textbf{q}$ increases, the acoustic and low-frequency breathing mode branches almost match being analogous with the low-frequency optical modes of bulk MoSe$_{2}$ and WSe$_{2}$ structures.
The LBM occurs only in bilayers with clean interface and are very sensitive to 
the vdW interactions and the relative orientation of the two monolayers.\cite{Lui2015} They are  expected to play an important role in the formation of  indirect interlayer excitons in vdW heterostructures.\cite{Gillen2018,Kunstmann2018} Furthermore, the understanding of this kind of phonon modes provides access to the intelayer interaction and to the type of the stacking arrangement as well. Note that the frequencies of the breathing phonon modes are larger than that of the shear phonon modes. The calculated phonon dispersion (see Fig.5) shows that the energy acoustic modes of the isolated constituents did not change mostly upon stacking indicating that the MoSe$_{2}$/WSe$_{2}$ heterobilayer exhibits  a thermal transport comparable to the ones of free-standing constituents. However, the thermal conductance could be deteriorated through acoustic-optic phonon scattering mechanisms involving low-frequency optical branches. The deviation of the flexural modes from characteristic quadratic  dependence and the rise of interlayer modes may be very useful to unravel further fundamental properties of the heterobilayer.

\section{Conclusions}
In conclusion, we have studied the structural  and vibrational properties  of MoSe$_{2}$/WSe$_{2}$ single layers and their van der Waals heterobilayer  MoSe$_{2}$/WSe$_{2}$ using advanced state-of-art first-principles calculations based on density functional perturbation theory and density functional theory. We find that the flexural modes have departed from quadratic wave vector characteristic of the isolated constituents. Furthermore, our calculations show the presence of interlayer shear and breathing phonon modes. The understanding of the behaviour of the vibrational properties allow us to controlling and monitoring the heat transfer in layered 2D materials. 

\section{Acknowledgments}
This work has been supported by the Algerian Ministry of High Education and Scientific Research under the PNE programme. The authors are grateful to J. Fern\'andez-Rossier for his outstanding help and valuable advice. F. M. thanks the hospitality of the Departamento de F\'isica Aplicada at the Universidad de Alicante. A. M.-S. acknowledges the Ram\'on y Cajal programme (grant RYC2018-024024-I; MINECO, Spain) and the Marie-Curie-COFUND program Nano TRAIN For Growth II (Grant Agreement 713640). $Ab ~ initio$ simulations were performed on the Tirant III cluster of the Servei d‘Informática of the University of Valencia and on the plateforme de Calcul intensif -HPC Ibnkhaldoun of the University of Biskra. 

%\bibliographystyle{apsrev4-1}
%\bibliography{references}{}

%\bibliographystyle{plain}
%\bibliography{references}
\end{document}